\documentclass[aps,pra,reprint,showpacs]{revtex4-1}
\usepackage{amsmath}
\usepackage{SIunits}
\usepackage{graphicx}

\begin{document}
%
%
%
\newcommand{\ac}[0]{\ensuremath{\hat{a}_{\mathrm{c}}}}
\newcommand{\adagc}[0]{\ensuremath{\hat{a}^{\dagger}_{\mathrm{c}}}}
\newcommand{\aR}[0]{\ensuremath{\hat{a}_{\mathrm{R}}}}
\newcommand{\aT}[0]{\ensuremath{\hat{a}_{\mathrm{T}}}}
\renewcommand{\b}[0]{\ensuremath{\hat{b}}}
\newcommand{\bdag}[0]{\ensuremath{\hat{b}^{\dagger}}}
\newcommand{\betaI}[0]{\ensuremath{\beta_\mathrm{I}}}
\newcommand{\betaR}[0]{\ensuremath{\beta_\mathrm{R}}}
\renewcommand{\c}[0]{\ensuremath{\hat{c}}}
\newcommand{\cdag}[0]{\ensuremath{\hat{c}^{\dagger}}}
\newcommand{\CorrMat}[0]{\ensuremath{\boldsymbol\gamma}}
\newcommand{\Deltacs}[0]{\ensuremath{\Delta_{\mathrm{cs}}}}
\newcommand{\Deltae}[0]{\ensuremath{\Delta_{\mathrm{e}}}}
\newcommand{\dens}[0]{\ensuremath{\hat{\rho}}}
\newcommand{\erfc}[0]{\ensuremath{\mathrm{erfc}}}
\newcommand{\gammapar}[0]{\ensuremath{\gamma_{\parallel}}}
\newcommand{\gammaperp}[0]{\ensuremath{\gamma_{\perp}}}
\newcommand{\gbar}[0]{\ensuremath{\bar{g}}}
\newcommand{\gens}[0]{\ensuremath{g_{\mathrm{ens}}}}
\renewcommand{\H}[0]{\ensuremath{\hat{H}}}
\renewcommand{\Im}[0]{\ensuremath{\mathrm{Im}}}
\newcommand{\kappac}[0]{\ensuremath{\kappa_{\mathrm{c}}}}
\newcommand{\mat}[1]{\ensuremath{\mathbf{#1}}}
\newcommand{\mean}[1]{\ensuremath{\langle#1\rangle}}
\newcommand{\omegac}[0]{\ensuremath{\omega_{\mathrm{c}}}}
\newcommand{\omegas}[0]{\ensuremath{\omega_{\mathrm{s}}}}
\newcommand{\pauli}[0]{\ensuremath{\hat{\sigma}}}
\newcommand{\Pa}[0]{\ensuremath{\hat{P}_{\mathrm{c}}}}
\renewcommand{\Re}[0]{\ensuremath{\mathrm{Re}}}
\renewcommand{\S}[0]{\ensuremath{\hat{S}}}
\newcommand{\Sminuseff}[0]{\ensuremath{\hat{S}_-^{\mathrm{eff}}}}
\newcommand{\Sxeff}[0]{\ensuremath{\hat{S}_x^{\mathrm{eff}}}}
\newcommand{\Syeff}[0]{\ensuremath{\hat{S}_y^{\mathrm{eff}}}}
\newcommand{\tildeac}[0]{\ensuremath{\tilde{a}_{\mathrm{c}}}}
\newcommand{\tildepauli}[0]{\ensuremath{\tilde{\sigma}}}
\newcommand{\Var}[0]{\ensuremath{\mathrm{Var}}}
\renewcommand{\vec}[1]{\ensuremath{\mathbf{#1}}}
\newcommand{\Xa}[0]{\ensuremath{\hat{X}_{\mathrm{c}}}}

\title{Dynamical evolution of an inverted spin ensemble in a cavity: \\
  Inhomogeneous broadening as a stabilizing mechanism}

\author{Brian Julsgaard}
\email{brianj@phys.au.dk}

\author{Klaus M{\o}lmer}
\affiliation{Lundbeck Foundation Theoretical Center for Quantum System
  Research, Department of Physics and Astronomy, Aarhus University, Ny
  Munkegade 120, DK-8000 Aarhus C, Denmark.}


\date{\today}

\begin{abstract}
  We study the evolution of an inverted spin ensemble coupled to a
  cavity. The inversion itself presents an inherent instability of the
  system; however, the inhomogeneous broadening of spin-resonance
  frequencies presents a stabilizing mechanism, and a stability
  criterion is derived. The detailed behavior of mean values and
  variances of the spin components and of the cavity field is
  accounted for under both stable and unstable conditions.
\end{abstract}


\maketitle

\section{Introduction}
\label{sec:introduction}
As an extension of traditional cavity-quantum-electro-dynamics
\cite{Kimble.PhysicaScripta.T76.127(1998)}, the resonant coupling of a
cavity to an ensemble of two-level systems has received considerable
interest for the past three decades. In particular, by the collective
effect of $N$ particles, the otherwise weak single-particle coupling
is enhanced by a factor of $\sqrt{N}$
\cite{Agarwal.PhysRevLett.53.1732(1984)}. In experiment, this has
allowed for reaching the collective strong-coupling regime in atomic
\cite{Kaluzny.PhysRevLett.51.1175(1983),
  Raizen.PhysRevLett.63.240(1989), Zhu.PhysRevLett.64.2499(1990)},
ionic \cite{Herskind.NaturePhys.5.494(2009)}, and solid-state
implementations \cite{Schuster.PhysRevLett.105.140501(2010),
  Kubo.PhysRevLett.105.140502(2010),
  Amsuss.PhysRevLett.107.060502(2011)}, typically materializing as a
normal-mode splitting of the coupled radiation-matter
system. Ensembles of electronic spins coupled to a micro-wave cavity
have recently been considered for quantum-memory purposes
\cite{Imamoglu.PhysRevLett.102.083602(2009),
  Wesenberg.PhysRevLett.103.070502(2009),
  Wu.PhysRevLett.105.140503(2010), Kubo.PhysRevLett.107.220501(2011),
  Yang.PhysRevA.84.010301R(2011)}. However, such ensembles usually
contain an inherent inhomogeneity of the spin transition frequencies,
which leads to dephasing of the stored information. While
spin-refocusing techniques reverse this process, it is necessary to
understand the dynamical effects and stability of an inverted ensemble
coupled to a cavity-field mode to benefit from such refocusing
processes. This is the topic of the present manuscript. We
demonstrate, in particular, that the inhomogeneity of the spin
transition frequencies is an advantage in the sense that it plays a
stabilizing role for an inverted ensemble.  Dynamical effects will be
examined for both mean values and second moments, and a stability
criterion for an inverted sample is derived.

Our analysis applies in general for any large collection of two-level
systems, but for convenience we shall use the terminology and notation
of ensembles of spin-$\frac{1}{2}$ particles. The paper is arranged as
follows: In Sec.~\ref{sec:equations-motion} the basic interaction and
decay mechanisms of the spin-cavity system is described, and the
dynamical evolution of the physical system is calculated with emphasis
on mean values in Sec.~\ref{sec:Evolution_mean_values} and on
second moments in Sec.~\ref{sec:Evolution_quadratic_moments}. A
few experimental diagnostics tools are suggested in
Sec.~\ref{sec:probing}, while a general discussion and conclusion of
the results are given in sections \ref{sec:Discussion}
and~\ref{sec:Conclusion}, respectively. Some mathematical details have
been deferred to appendix~\ref{app:Discretization}.

\section{Equations of motion}
\label{sec:equations-motion}
\begin{figure}[t]
  \centering
  \includegraphics{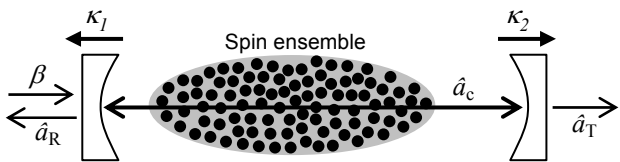}
  \caption{The physical setup under consideration. A spin ensemble is
    coupled to a cavity field $\ac$, which is subjected to decay
    through the two mirrors with field-decay rates $\kappa_1$ and
    $\kappa_2$. For diagnostics purposes an external driving field
    $\beta$ may be applied giving rise to reflected and transmitted
    fields, $\aR$ and $\aT$, as discussed in Sec.~\ref{sec:probing}.}
\label{fig:CavitySetup}
\end{figure}

We consider an ensemble of $N$ spins coupled to a single-mode cavity
field, $\ac$, as shown in Fig.~\ref{fig:CavitySetup}. The resonance
frequency, $\omega_j$, of each spin is assumed to be inhomogeneously
broadened around a central frequency, $\omegas$, and the coupling
strength, $g_j$, between individual spins and the cavity field may
also vary.  An external field, $\beta$, may be used to drive the
cavity field through the left-most mirror with field-decay rate,
$\kappa_1$ (in the present manuscript this driving field is only used
for diagnostics and otherwise left at zero).  In the frame rotating at
the central spin frequency, $\omegas$, the Hamiltonian can be
expressed as:
\begin{align}
\notag
  \H &= \hbar\Deltacs\adagc\ac + \frac{\hbar}{2}\sum_{j=1}^N\Delta_j\pauli_z^{(j)}
      + i\hbar\sqrt{2\kappa_1}(\beta\adagc - \beta^*\ac) \\
    &\quad + \hbar \sum_{j=1}^N g_j(\pauli_+^{(j)}\ac + \pauli_-^{(j)}\adagc),
\label{eq:Hamiltonian_working}
\end{align}
where $\Deltacs = \omegac - \omegas$ is the detuning of the cavity
resonance frequency $\omegac$ from $\omegas$, and $\Delta_j =
\omega_j-\omegas$. The Pauli operators $\pauli_k^{(j)}$ with $k =
-,+,z$ are used to model the $j$'th spin. The $c$-number, $\beta$,
represents an external coherent-state driving field and is normalized
such that $|\beta|^2$ is the incoming number of photons per second.

Decay mechanisms are taken into account in the Markovian approximation
of memoryless reservoirs: Cavity leakage is parametrized by the total
field-decay rate $\kappa = \kappa_1 + \kappa_2$, and the dephasing
rate $\gammaperp = \frac{1}{\tau}$ represents the loss of coherence at
the level of individual spins with a characteristic coherence time
$\tau$.

By defining $\b = \frac{1}{\gens}\sum_{j=1}^N g_j\pauli_-^{(j)}$ and
$\bdag = \frac{1}{\gens}\sum_{j=1}^N g_j\pauli_+^{(j)}$, where the
ensemble-coupling constant $\gens$ is given by $\gens^2 = \sum_{j=1}^N
g_j^2$, the interaction part of the
Hamiltonian~(\ref{eq:Hamiltonian_working}) can be written as
$\hbar\gens(\ac\bdag + \adagc\b)$. This becomes particularly useful
when essentially all spins are in the ground state ($\pauli_z^{(j)}
\approx -1$ and $[\b,\bdag] \approx 1$ in which case the spin system
can be represented by a harmonic oscillator - the so-called
Holstein-Primakoff approximation
\cite{Holstein.PhysRev.58.1098(1940)}). The resulting formal
equivalence between quantized fields and collective spin degrees of
freedom, and their coupling strength which is collectively enhanced by
a factor of $\sqrt{N}$, have paved the way for using spin ensembles
for quantum information purposes
\cite{Imamoglu.PhysRevLett.102.083602(2009),
  Wesenberg.PhysRevLett.103.070502(2009),
  Wu.PhysRevLett.105.140503(2010), Kubo.PhysRevLett.107.220501(2011),
  Yang.PhysRevA.84.010301R(2011)}. The Holstein-Primakoff Hamiltonian
is quadratic in the oscillator quadrature operators, which implies
that first and second moments of those operators are described by a
closed set of equations, also in the presence of inhomogeneous
coupling \cite{Madsen.PhysRevA.70.052324(2004)} and broadening. The
influence of inhomogeneous broadening on a spin-cavity system has been
studied previously under the Holstein-Primakoff approximation for
ensembles essentially in the ground state
\cite{Houdre.PhysRevA.53.2711(1996), Kurucz.PhysRevA.83.053852(2011),
  Diniz.PhysRevA.84.063810(2011),
  Sandner.PhysRevA.85.053806(2012)}. For an inverted ensemble
containing 10 two-level systems the evolution of the mean values was
studied phenomenologically in
Ref.~\cite{Temnov.PhysRevLett.95.243602(2005)}. The present manuscript
is focused on large inverted ensembles, in which case the convenient
description of both first and second moments under the
Holstein-Primakoff approximation is possible.

\section{Dynamical evolution of an inverted medium inside a cavity:
  Mean values}
\label{sec:Evolution_mean_values}
The present section considers the evolution of mean values of the
cavity field and the spin components for an inverted spin state. The
calculations assume a resonant coupling between the cavity and the
spins, $\Deltacs = 0$, in which case the effects under study are
strongest.
\subsection{A stability criterion using the effective cooperativity parameter}
Consider the following mean value equations, which have been derived
under the Holstein-Primakoff approximation ($\pauli_z^{(j)} \approx
1$) in absence of external driving:
\begin{align}
\label{eq:ddt_ac_no_driving}
  \frac{\partial\mean{\ac}}{\partial t} &= -(\kappa+i\Deltacs)\mean{\ac}
    -i\sum_{j=1}^N g_j\mean{\pauli_-^{(j)}}, \\
\label{eq:ddt_sigmaMinus_inverted}
  \frac{\partial\mean{\pauli_-^{(j)}}}{\partial t} &=
    -(\gammaperp+i\Delta_j)\mean{\pauli_-^{(j)}} + ig_j\mean{\ac}.
\end{align}
We note that if $\mean{\ac}$ is real and positive, the second term of
Eq.~(\ref{eq:ddt_sigmaMinus_inverted}) will drive
$\mean{\pauli_-^{(j)}}$ toward positive imaginary values. In turn,
the second term of Eq.~(\ref{eq:ddt_ac_no_driving}) will drive
$\mean{\ac}$ further along the positive real axis, and the physical
system is thus unstable due to the gain provided by the inverted
sample. This scenario resembles to a large extent a laser, and normal
laser operation is initiated when the gain medium is able to balance
the optical losses of the cavity; however, the case under study here
differs from normal laser operation by the fact that the inverted spin
medium behaves coherently. Accordingly, a large cavity loss (i.e.~a
large $\kappa$) is not the only way to counter-act the inherent
instability, but dephasing due to inhomogeneous broadening will also
contribute.

In analogy to threshold conditions for normal laser operation, a
stability criterion can be derived for our spin-cavity system by
searching for a critical cavity-coupling parameter, $\kappac$, which
allows for a non-zero steady-state solution for $\mean{\ac}$ and
$\mean{\pauli_j^{(j)}}$. Then increasing (decreasing) solutions versus
time are expected when $\kappa < \kappac$ ($\kappa > \kappac$).  To
this end, consider first Eq.~(\ref{eq:ddt_sigmaMinus_inverted}) in
steady state: $\mean{\pauli_-^{(j)}} = \frac{ig_j\mean{\ac}}
{\gammaperp+i\Delta_j}$, which inserted into
Eq.~(\ref{eq:ddt_ac_no_driving}) in steady state leads to:
$(\kappa+i\Deltacs)\mean{\ac} = \sum_j\frac{g_j^2} {\gammaperp +
  i\Delta_j}\mean{\ac}$. We shall restrict ourselves to inhomogeneous
broadening with $\Delta_j$ distributed symmetrically around zero, in
which case $\Deltacs = 0$ is indeed the relevant choice. The above
equation can be satisfied for a non-zero $\mean{\ac}$ provided that
$\kappa$ attains the critical value:
\begin{equation}
\label{eq:Def_kappac_and_Gamma}
  \kappac = \gens^2\int_{-\infty}^{\infty}\frac{f(\Delta)d\Delta}
     {\gammaperp + i\Delta} \equiv \frac{\gens^2}{\Gamma},
\end{equation}
where we assumed the distributions of $g_j$ and $\Delta_j$ to be
uncorrelated. Furthermore, the continuum limit was taken by using the
spin-resonance-frequency distribution $f(\Delta)$ normalized such that
$\int_{-\infty}^{\infty}f(\Delta)d\Delta = 1$. The
\emph{characteristic width} $\Gamma$ of the inhomogeneous distribution
was implicitly defined, and the requirement of $\kappa > \kappac$ for
stability can be reformulated in terms of the \emph{effective}
cooperativity parameter, $C$:
\begin{equation}
\label{eq:Stability_condition}
  C = \frac{\gens^2}{\kappa\Gamma} < 1.
\end{equation}

\subsection{Homogeneous broadening}
Even though the main focus of this paper is inhomogeneous broadening,
it is convenient to know the effects of homogeneous broadening for
comparison. From Eq.~(\ref{eq:Def_kappac_and_Gamma}) it follows
immediately that $\Gamma = \gammaperp$ in this case ($f(\Delta)$ is a
$\delta$-function). In fact, Eqs.~(\ref{eq:ddt_ac_no_driving})
and~(\ref{eq:ddt_sigmaMinus_inverted}) can be reformulated in terms of
the \emph{effective} spin component $\Sminuseff = \sum_{j=1}^N
\frac{g_j}{\gbar}\pauli_-^{(j)}$, where $\gbar^2 =
\sum_{j=1}^Ng_j^2/N$, and the inverted spin-state problem is only
two-dimensional:
\begin{equation}
\label{eq:ddt_ac_and_Sminuseff_inv_noDrive}
  \frac{\partial}{\partial t}
  \begin{bmatrix}
    \mean{\ac} \\ \mean{\Sminuseff}
  \end{bmatrix}
  =
  \begin{bmatrix}
    -(\kappa+i\Deltacs) & -i\gbar  \\
    i\gbar N & -\gammaperp
  \end{bmatrix}
  \begin{bmatrix}
    \mean{\ac} \\ \mean{\Sminuseff}
  \end{bmatrix}.
\end{equation}
On resonance, $\Deltacs = 0$, the eigenvalues of this linear set of
equations are:
\begin{equation}
\label{eq:lambda_mean_value_hom_broad}
  \lambda_{\pm} = -\frac{\kappa+\Gamma}{2}\left(1 \mp
    \sqrt{1+\frac{4(C-1)\kappa\Gamma}{(\kappa+\Gamma)^2}}\right).
\end{equation}
Clearly, when $C<1$ both eigenvalues are negative and the inverted
spin state with $\mean{\ac} = \mean{\Sminuseff} = 0$ is a stable
solution.

\subsection{Inhomogeneous broadening}
In order to examine the dynamical evolution of the spin-cavity system
with analytical methods in the case of inhomogeneous broadening, it is
convenient to treat Eqs.~(\ref{eq:ddt_ac_no_driving})
and~(\ref{eq:ddt_sigmaMinus_inverted}) in Fourier space. In order to
handle also exponentially increasing solutions, we re-write the
dynamical variables as $\mean{\ac} = \mean{\tildeac}e^{\eta t}$ and
$\mean{\pauli_-^{(j)}} = \mean{\tildepauli_-^{(j)}}e^{\eta t}$. Assume
also that $\mean{\ac} = \mean{\pauli_-^{(j)}} = 0$ when $t < 0$, which
indeed presents a mathematical solution to the differential
equations. Then, at $t=0$ we change abruptly the cavity-field mean
value $\mean{\ac} \rightarrow \alpha$ and study the subsequent
dynamics. This scenario is governed by a modified version of
Eqs.~(\ref{eq:ddt_ac_no_driving})
and~(\ref{eq:ddt_sigmaMinus_inverted}) taken at resonance, $\Deltacs =
0$:
\begin{align}
\label{eq:ddt_ac_with_delta_kick}
  \frac{\partial\mean{\tildeac}}{\partial t} &= \alpha\delta(t)
    -(\kappa+\eta)\mean{\tildeac} -i\sum_{j=1}^N g_j\mean{\tildepauli_-^{(j)}},\\
\label{eq:ddt_sigmaMinus_inverted_special}
  \frac{\partial\mean{\tildepauli_-^{(j)}}}{\partial t} &=
    -(\gammaperp+\eta+i\Delta_j)\mean{\tildepauli_-^{(j)}} + ig_j\mean{\tildeac}.
\end{align}
The latter of these can be integrated formally:
$\mean{\tildepauli_-^{(j)}(t)} = ig_j\int_0^t e^{-(\gammaperp + \eta +
  i\Delta_j)(t-t')}\mean{\tildeac(t')} dt'$, which in turn can be
inserted into Eq.~(\ref{eq:ddt_ac_with_delta_kick}):
\begin{equation}
  \frac{\partial\mean{\tildeac}}{\partial t} = \alpha\delta(t)
  -(\kappa+\eta)\mean{\tildeac} + \int_0^t \tilde{K}(t-t')\mean{\tildeac(t')},
\end{equation}
where $\tilde{K}(t) = \sum_{j=1}^N g_j^2
e^{-(\gammaperp+\eta+i\Delta_j)t}$. Now, by defining the positive-time
version of $\tilde{K}$ by $\tilde{K}^+(t) = \tilde{K}(t)\cdot
\theta(t)$, where $\theta(t)$ is the Heaviside step function, and by
remembering that $\mean{\tildeac(t')} = 0$ when $t'<0$, the above
integration can be extended to plus/minus infinity. Using the Fourier
transform, $\mean{\tildeac(\omega)} = \int_{-\infty}^{\infty}
\mean{\tildeac(t)} e^{i\omega t}dt$ and $\mean{\tildeac(t)} =
\frac{1}{2\pi} \int_{-\infty}^{\infty} \mean{\tildeac(\omega)}
e^{-i\omega t}d\omega$, we find:
\begin{equation}
\label{eq:tildeac_of_omega}
  \mean{\tildeac(\omega)} = \frac{\alpha}{\kappa+\eta -i\omega -
    \tilde{K}^+(\omega)}.
\end{equation}
The Fourier transform $\tilde{K}^+(\omega)$ can be expressed in the
continuum limit as:
\begin{equation}
\label{eq:Formula_Ktilde_omega}
  \tilde{K}^+(\omega) = -i\gens^2\int_{-\infty}^{\infty}\frac{f(\Delta)d\Delta}
    {\Delta-\omega-i(\gammaperp+\eta)}.
\end{equation}
\subsubsection{Lorentzian broadening}
\begin{figure}[t]
  \centering
  \includegraphics{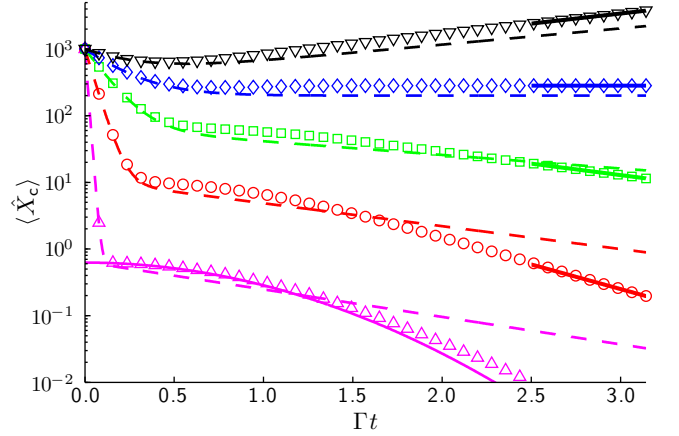}
  \caption{(Color online) Cavity-field decay versus time when the
    inhomogeneous broadening is Lorentzian [dashed lines, from
    Eq.~(\ref{eq:ac_delta_pulse_Lorentzian})] and Gaussian [symbols,
    numerical simulation]. In all cases, $\gens = 2\Gamma$ and
    $\gammaperp = 0$, while $\kappa$ is varied such that $C$ attains
    the values of 0.05 (magenta tip-up triangles), 0.2 (red circles),
    0.5 (green squares), 1 (blue diamonds), and 2 (black tip-down
    triangles). The vertical axis contains $\mean{\Xa} = (\mean{\ac} +
    \mean{\adagc})/\sqrt{2}$. In the range $\Gamma t \ge 2.5$ the
    slope of the Gaussian-broadening decay curves are compared to a
    numerically determined value (solid lines) as discussed in the
    text prior to Eq.~(\ref{eq:lambda_gaussian_long_time_approx}). The
    solid curve through the magenta tip-up triangles is given by
    Eq.~(\ref{eq:ac_kick_decay_low_coupling}).}
\label{fig:GaussLorentzComparison}
\end{figure}
For a Lorentzian broadened spin ensemble with $f(\Delta) =
\frac{w/2\pi}{\Delta^2 + w^2/4}$, where $w$ is the FWHM (full width at
half maximum), the characteristic width of
Eq.~(\ref{eq:Def_kappac_and_Gamma}) becomes: $\Gamma = \frac{w}{2} +
\gammaperp$. Furthermore, Eq.~(\ref{eq:Formula_Ktilde_omega}) can be
written (using the residue theorem):
\begin{equation}
  \tilde{K}^+(\omega) = \frac{\gens^2}{\Gamma+\eta-i\omega},
\end{equation}
and inserting this result into Eq.~(\ref{eq:tildeac_of_omega}) leads
to:
\begin{equation}
\label{eq:tildeac_omega_Lorentzian}
  \mean{\tildeac(\omega)} = \frac{\alpha(i\omega - \Gamma - \eta)}
    {(\omega -i[\lambda_+-\eta])(\omega -i[\lambda_--\eta])},
\end{equation}
where $\lambda_{\pm}$ are the solutions given in
Eq.~(\ref{eq:lambda_mean_value_hom_broad}). The inverse Fourier
transform is now invoked, leading to ($t>0$):
\begin{equation}
\label{eq:ac_delta_pulse_Lorentzian}
  \mean{\ac(t)} = \alpha\frac{(\lambda_++\Gamma)e^{\lambda_+t} -
    (\lambda_-+\Gamma)e^{\lambda_-t}}{\lambda_+ - \lambda_-},
\end{equation}
and $\mean{\ac(t)} = 0$ when $t<0$. This expression is independent of
$\eta$ as it should be; however, for the Fourier transform
$\mean{\tildeac(\omega)}$ to exist, the condition $\eta >
\lambda_+$ must be fulfilled, which in fact also ensures that
both poles in Eq.~(\ref{eq:tildeac_omega_Lorentzian}) reside in the
lower complex half-plane.
\subsubsection{Gaussian broadening}
For a Gaussian broadened spin ensemble with $f(\Delta) =
\frac{1}{\sqrt{2\pi}\sigma_{\Delta}}e^{-\Delta^2/2\sigma_{\Delta}^2}$,
where $\sigma_{\Delta}$ is the standard deviation of the distribution
(connected to the FWHM by $w = \sigma_{\Delta}\sqrt{8\ln(2)}$) the
characteristic width~(\ref{eq:Def_kappac_and_Gamma}) is given by:
$\Gamma = \sqrt{\frac{2}{\pi}}\frac{\sigma_{\Delta}}{w(z)}$, where the
\emph{complex error function} is given by $w(z) = e^{-z^2}\erfc(-iz)$
with $\erfc(\cdot)$ being the complementary error function
\cite{AbramowitzStegun} and $z =
\frac{i\gammaperp}{\sqrt{2}\sigma_{\Delta}}$.
Equation~(\ref{eq:Formula_Ktilde_omega}) reads in this case:
\begin{equation}
  \tilde{K}^+(\omega) = \sqrt{\frac{\pi}{2}}\frac{\gens^2}{\sigma_{\Delta}}
   w(\tilde{z}),
\end{equation}
where $\tilde{z} = \frac{\omega + i (\gammaperp+\eta)}
{\sqrt{2}\sigma_{\Delta}}$. It is not possible to write a general
analytic expression for the inverse transform $\mean{\ac(t)}$;
however, we shall calculate $\mean{\ac(t)}$ by numerical integration
of Eq.~(\ref{eq:dydt=My}) with a real, non-zero $X_{\mathrm{c}} =
\sqrt{2}\alpha$ as initial condition at $t=0$, and limiting cases will
be compared to analytical estimates. Such numerical simulations are
shown (with symbols) in Fig.~\ref{fig:GaussLorentzComparison} for
various values of the effective cooperativity parameter $C$, and
comparison to the Lorentzian-broadened case is made (by dashed lines,
maintaining $\kappa$ and $\Gamma$). The following points can be noted:
(I) the initial decay seems similar for Lorentzian and Gaussian
broadening, (II) in the long-time limit for Gaussian broadening the
decay seems to be exponential, and (III) when the coupling is weak ($C
\ll 1$) the curves for Gaussian broadening appear to have a
significant quadratic content when plotted on the logarithmic vertical
scale.

The single-exponential parts of the decay curves (with rate $\lambda$)
correspond to the poles of Eq.~(\ref{eq:tildeac_of_omega}),
i.e.~solutions ($\omega = i\lambda$) to the equation $\kappa-i\omega =
\sqrt{\frac{\pi}{2}}\frac{\gens^2}{\sigma_{\Delta}}w(\tilde{z})$
taking $\eta = 0$. Provided that $|\lambda| \gg \sigma_{\Delta}$ for
the initial fast decay, we take advantage of the series expansion,
$w(z) \approx \frac{i}{\sqrt{\pi}z}$ when $|z| \gg 1$, and reach the
condition: $(\lambda+\kappa)(\lambda+\gammaperp) = \gens^2$. This is
exactly the eigen-value equation for the \emph{homogeneous} system of
equations~(\ref{eq:ddt_ac_and_Sminuseff_inv_noDrive}), and the
solution $\lambda$ is equal to $\lambda_-$ in
Eq.~(\ref{eq:lambda_mean_value_hom_broad}) with $\Gamma =
\gammaperp$. From a physical perspective, the narrow feature of the
Gaussian broadening cannot be resolved on the initial fast time
scales. The long-time limit of the decay for Gaussian broadening is
compared in Fig.~\ref{fig:GaussLorentzComparison} by solid lines to
the rate $\lambda$ found by locating numerically another pole of
Eq.~(\ref{eq:tildeac_of_omega}). In the vicinity of the stability
threshold, $\kappa \approx \kappac$ such that $|\lambda| \ll
\sigma_{\Delta}$, the value of $\lambda$ can be approximated by using
the series expansion, $w(z) \approx 1+\frac{2iz}{\sqrt{\pi}} - z^2$
when $|z| \ll 1$, leading to:
\begin{equation}
\label{eq:lambda_gaussian_long_time_approx}
  \lambda \approx \frac{\kappac-\kappa}{1+\frac{\gens^2}{\sigma_{\Delta}^2}}
   + \sqrt{\frac{\pi}{8}}\frac{\gens^2}{\sigma_{\Delta}^3}
   \frac{(\kappac-\kappa)^2}{\left(1+\frac{\gens^2}{\sigma_{\Delta}^2}\right)^3}.
\end{equation}
Finally, the weak-coupling limit, $C \ll 1$, can be calculated
directly from Eqs.~(\ref{eq:ddt_ac_with_delta_kick})
and~(\ref{eq:ddt_sigmaMinus_inverted_special}), provided that $\kappa$
is faster than the remaining dynamical processes. In a first
approximation, $\mean{\ac(t)} = \alpha e^{-\kappa t}$, since the spins
will contribute little due to the low coupling. Secondly, during the
initial decay of the cavity field, each spin component acquires a
small value: $\mean{\pauli_-^{(j)}} = \frac{ig_j\alpha}{\kappa}$,
which is derived by integrating the second term of
Eq.~(\ref{eq:ddt_sigmaMinus_inverted_special}); the first term can be
neglected on this fast time scale. Thirdly, after the initial cavity
decay, the spins evolve freely due to the low coupling:
$\frac{\partial}{\partial t}\mean{\pauli_-^{(j)}} =
-(\gammaperp+i\Delta_j)\mean{\pauli_-^{(j)}}$, and the cavity field
follows the spins adiabatically in this regime:
\begin{equation}
\label{eq:ac_kick_decay_low_coupling}
  \mean{\ac(t)} \approx -\frac{i}{\kappa}\sum_{j=1}^Ng_j\mean{\pauli_-^{(j)}}
    = \frac{\alpha\gens^2}{\kappa^2}e^{-\frac{1}{2}\sigma_{\Delta}^2 t^2-\gammaperp t},
\end{equation}
where the continuum limit of the inhomogeneous frequency distribution
was taken in the last step. Alternatively, when $C \ll 1$,
Eq.~(\ref{eq:tildeac_of_omega}) can be approximated:
$\mean{\tildeac(\omega)} \approx \frac{\alpha}{\kappa-i\omega} +
\frac{\alpha\gens^2} {(\kappa-i\omega)^2} \sqrt{\frac{\pi}{2}}
\frac{w(z)}{\sigma_{\Delta}} \approx \frac{\alpha}{\kappa-i\omega} +
\frac{\alpha\gens^2} {\kappa^2} \sqrt{\frac{\pi}{2}}
\frac{w(z)}{\sigma_{\Delta}}$, where the second step considers only
the low-frequency parts of the second term ($w(z)$ varies on the
frequency scale of $\sigma_{\Delta} \ll \kappa$). The inverse Fourier
transform of this approximated $\mean{\tildeac(\omega)}$ is $\alpha
e^{-\kappa t}$ plus the term found in
Eq.~(\ref{eq:ac_kick_decay_low_coupling}). The lower curve (magenta
tip-up triangles, $C = 0.05$) in Fig.~\ref{fig:GaussLorentzComparison}
follows Eq.~(\ref{eq:ac_kick_decay_low_coupling}) to a large extent.
\section{Dynamical evolution of an inverted medium inside a cavity:
  Quadratic moments}
\label{sec:Evolution_quadratic_moments}
The calculation of the dynamical evolution of mean values in the
preceding section presents one of the main results of the present
manuscript. However, we wish to back up these mean-field results by a
calculation of second moments --- a mean-value stabilized
spin-cavity system would be of less relevance if e.g.~the variance of
the spin components and the cavity field increased without
limits. Such an unlimited increase will also render the spin-cavity
system inapplicable for quantum-memory purposes.

The case of homogeneous broadening is treated analytically while
inhomogeneous broadening requires numerical treatment. In any case,
the calculations follow the general procedure outlined in appendix
~\ref{app:Discretization}.
\subsection{Homogeneous broadening}

Assume that $\Delta_j = 0$ and $g_j = g$ for all spins. For an
inverted spin sample on resonance with mean values $X_{\mathrm{c}} =
P_{\mathrm{c}} = S_x = S_y = 0$, $S_z = N$, and $\Deltacs = 0$, we
introduce the vector of second moments, $\vec{x} =
[\mean{\delta\Xa^2}$ $\mean{\delta\Pa^2}$ $\mean{\delta\S_x^2}$
$\mean{\delta\S_y^2}$ $\mean{\delta\S_x\delta\Pa}$
$\mean{\delta\S_y\delta\Xa}]^{\mathrm{T}}$. Following
Eq.~(\ref{eq:ddt_CorrMat}) in the appendix, they obey the following
set of coupled equations, $\frac{\partial \vec{x}}{\partial t} =
\mat{Q}\vec{x} + \vec{r}$, where
\begin{widetext}
\begin{equation}
  \vec{Q} =
  \begin{bmatrix}
    -2\kappa & 0 & 0 & 0 & 0 & -\sqrt{2}g \\
    0 & -2\kappa & 0 & 0 & -\sqrt{2}g & 0 \\
    0 & 0 & -2\gammaperp & 0 & -2\sqrt{2}gN & 0 \\
    0 & 0 & 0 & -2\gammaperp & 0 & -2\sqrt{2}gN \\
    0 & -\sqrt{2}gN & -\frac{g}{\sqrt{2}} & 0 & -(\kappa+\gammaperp) & 0 \\
    -\sqrt{2}gN & 0 & 0 & -\frac{g}{\sqrt{2}} & 0 & -(\kappa+\gammaperp)
  \end{bmatrix}, \quad
  \vec{r} =
  \begin{bmatrix}
    \kappa \\ \kappa \\ 2\gammaperp N \\ 2\gammaperp N \\ 0 \\ 0
  \end{bmatrix}.
\end{equation}
\end{widetext}
In fact, an inhomogeneous distribution of the coupling constants,
$g_j$, can be incorporated in the above equations by merely replacing
$\S_x \rightarrow \Sxeff$, $\S_y \rightarrow \Syeff$, and
$g\rightarrow \gbar$.  The matrix $\mat{Q}$ has three
doubly-degenerate eigenvalues. Two of these are given by $\lambda =
2\lambda_{\pm}$, i.e.~by twice the values found in
Eq.~(\ref{eq:lambda_mean_value_hom_broad}), and the third one is
$\lambda = -(\kappa+\gammaperp)$. Hence, the same condition $C < 1$,
ensures that both the first and second moments are stable and converge
to their steady-state values. Solving $\frac{\partial
  \vec{x}}{\partial t} =\mat{Q}\vec{x}+\vec{r} = 0$, these read:
\begin{equation}
\label{eq:Variances_SS}
  \begin{split}
    \mean{\delta\Xa^2} &= \mean{\delta\Pa^2} = \frac{1}{2}\cdot
     \frac{1-C\frac{\kappa-\Gamma}{\kappa+\Gamma}}{1-C}, \\
   \mean{\delta\hat{S}_x^{\mathrm{eff}\,2}} &=
     \mean{\delta\hat{S}_y^{\mathrm{eff}\,2}} = N\cdot
     \frac{1+C\frac{\kappa-\Gamma}{\kappa+\Gamma}}{1-C}, \\
   \mean{\delta\S_x^{\mathrm{eff}}\delta\Pa} &=
     \mean{\delta\S_y^{\mathrm{eff}}\delta\Xa} =
     -\sqrt{\frac{N}{2}}\frac{2\gens}{(\kappa+\Gamma)(1-C)},
  \end{split}
\end{equation}
where $\Gamma = \gammaperp$ for homogeneous broadening. We note that
the levels of $\frac{1}{2}$ and $N$ correspond to the variance of the
minimum-uncertainty states for the cavity field and the collective
spin, respectively. When approaching the stability point $C\rightarrow
1$ from below, the variances diverge.

\subsection{Inhomogeneous broadening}
\label{sec:QuadMoments_inhomogeneous}
\begin{figure}[t]
  \centering
  \includegraphics{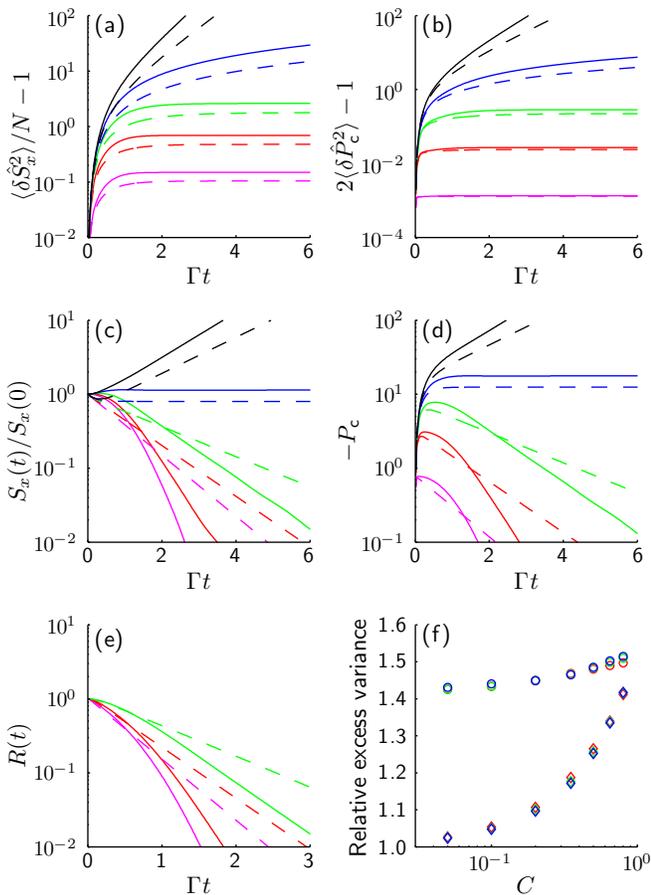}
  \caption{(Color online) Panels (a-e) show various dynamical
    parameters versus time with fixed $\gens = 2\Gamma$ and varying
    $\kappa$, such that $C$ attains the values $0.05$ (magenta), $0.2$
    (red), $0.5$ (green), $1$ (blue), and $2$ (black), represented in
    the order from the the lower to the upper sets of dashed and solid
    curves in all plots.  Solid lines have been calculated for
    Gaussian inhomogeneous broadening ($\gammaperp = 0$), while the
    dashed lines correspond to both Lorentzian broadening ($\gammaperp
    = 0$) and homogeneous broadening ($\gammaperp = \Gamma$). (a)
    Excess variance of $\S_x$ relative to the coherent-state value of
    $N$. (b) Excess variance of $\Pa$ relative to the coherent-state
    value of $\frac{1}{2}$. (c) The mean value of $S_x$ normalized to
    $S_x(t=0)$. (d) The mean value of $\Pa$ in arbitrary units. (e)
    The relative deviation of $\mean{\delta\S_x^2}$ from its
    steady-state value, $R(t)$, see the text for details. (f) The
    Gaussian-broadened steady-state excess variance
    $[\mean{\delta\S_x^2(\infty)}/N-1]$ (circles) or
    $[2\mean{\delta\Pa^2(\infty)}-1]$ (diamonds), relative to the
    corresponding values from Eq.~(\ref{eq:Variances_SS}). The results
    are very similar for the different coupling strengths $\gens =
    3\Gamma$ (red), $\gens = 4\Gamma$ (green), and $\gens = 5\Gamma$
    (blue).}
\label{fig:MomentEvolution}
\end{figure}
For the case of inhomogeneous broadening we use numerical simulation
of Eqs.~(\ref{eq:dydt=My}) and~(\ref{eq:ddt_CorrMat}) for calculating
the dynamical evolution. We use only a single value for $g_m$ but
choose either a Lorentzian or Gaussian shaped distribution of
$\Delta_m$. As a starting point, all spins are prepared in the
inverted coherent state being slightly displaced: $S_z^{(m)} = N_m$,
$S_y^{(m)} = 0$, and $S_x^{(m)} = \theta S_z^{(m)}$ where $\theta =
10^{-3}$, and the cavity is prepared in the vacuum state. Leaving the
spin-cavity system to evolve from this initial state, we study as
function of time a representative set of mean values and variances:
$S_x = \sum_{m=1}^M S_x^{(m)}$, $P_{\mathrm{c}}$, $\mean{\delta\S_x^2}
= \sum_{m,n=1}^M \mean{\delta\S_x^{(m)}\delta\S_x^{(n)}}$, and
$\mean{\delta\Pa^2}$, the results have been plotted in
Fig.~\ref{fig:MomentEvolution}.

Panels (a) and (b) of this figure show how the variances,
$\mean{\delta\S_x^2}$ and $\mean{\delta\Pa^2}$, increase from their
initial values of $N$ and $\frac{1}{2}$, respectively. As can be seen,
in the stable region with $C < 1$ these variances converge to a
steady-state value while for $C \ge 1$ the curves increase without
limits. The solid lines correspond to a Gaussian distribution while
the dashed lines correspond to a homogeneously broadened sample with
$\Gamma = \gammaperp$, which coincides with the simulations for a
Lorentzian broadened sample with $\gammaperp = 0$ and $\Gamma =
\frac{w}{2}$. At the same time, the mean values of $\S_x$ and $\Pa$
have been plotted in panel (c) and (d), which confirm that solutions
increase or decrease versus time when $C > 1$ or $C < 1$,
respectively. We note that the features and interpretation of these
graphs are very similar to those of
Fig.~\ref{fig:GaussLorentzComparison}; only the initial state is
different in the two figures. In order to show the time scale of the
dynamical evolution of variances, the deviation of
$\mean{\delta\S_x^2(t)}$ from its asymptotic value of
$\mean{\delta\S_x^2(\infty)}$ has been shown in panel (e) relative to
the entire dynamical range, i.e.~the vertical scale is the ratio:
$R(t) = \frac{\mean{\delta\S_x^2(\infty)} - \mean{\delta\S_x^2(t)}}
{\mean{\delta\S_x^2(\infty)} - \mean{\delta\S_x^2(0)}}$. Noting that
in panel (e) the horizontal axis spans only half the time as compared
to panel (c), it can be seen that the variance $\mean{\delta\S_x^2}$
approaches its asymptotic value approximately twice as fast as the
decay of the mean value $S_x$ toward zero. This is no surprise for the
homogeneous or Lorentzian case since we already observed that the
three characteristic eigenvalues of the problem, $2\lambda_+$,
$2\lambda_-$, and $\lambda_++\lambda_-$, relate closely to the eigen
values of the mean value
equation~(\ref{eq:lambda_mean_value_hom_broad}). The similarity of the
solid lines in panels (c) and (e) demonstrate that this holds
qualitatively also for the case of Gaussian broadening. Finally, it
can be observed from panels (a) and (b) that in the case of Gaussian
broadening (solid lines), the variances $\mean{\delta\S_x^2}$ and
$\mean{\delta\Pa^2}$ converge to values which are slightly higher than
those give by Eq.~(\ref{eq:Variances_SS}) when simply inserting the
corresponding values for $\kappa$, $\Gamma$, and $C$ (dashed
lines). The ratio of solid-to-dashed lines in panels (a) and (b) have
been shown in panel (f) with circles and diamonds, respectively, and
varying values for the ratio $\gens/\Gamma$ have been examined. We
conclude that Eq.~(\ref{eq:Variances_SS}) is not accurate for a
Gaussian inhomogeneous distribution although the qualitative features
remain.
\section{External probing of the spin sample}
\label{sec:probing}
The linear response of the spin ensemble can be probed by applying a
weak, external field $\beta$ and measuring the reflected or
transmitted field as depicted in Fig.~\ref{fig:CavitySetup}. Such a
measurement enables the determination of $C$ from a non-inverted
sample and also allows for assessing the efficiency of the
spin-inversion process. Assuming the cavity to be resonant with the
spins, $\Deltacs = 0$, the following mean-value equations are valid
under the Holstein-Primakoff approximation:
\begin{align}
  \frac{\partial\mean{\ac}}{\partial t} &= -\kappa\mean{\ac}
    -i\sum_{j=1}^N g_j\mean{\pauli_-^{(j)}} + \sqrt{2\kappa_1}\beta \\
  \frac{\partial\mean{\pauli_-^{(j)}}}{\partial t} &=
    -(\gammaperp+i\Delta_j)\mean{\pauli_-^{(j)}} + ipg_j\mean{\ac},
\end{align}
where $p = 1$ for an inverted sample and $p=-1$ for a non-inverted
sample. By applying a monochromatic external field, $\beta(t) =
\beta_0 e^{-i\Deltae t}$, the cavity-field mean value can be shown to
be:
\begin{equation}
  \mean{\ac(t)} = \frac{\sqrt{2\kappa_1}\beta(t)}
  {\kappa-i\Deltae - p\gens^2\int_{-\infty}^{\infty}
    \frac{f(\Delta)d\Delta}{\gammaperp+i(\Delta-\Deltae)}}.
\end{equation}
In fact, this is a particular solution to the differential equation
and we assume that the homogeneous solution has relaxed to zero; this
relaxation process was the topic of
Sec.~\ref{sec:Evolution_mean_values}, and for an inverted sample
($p=1$) the calculations only make sense if the stability criterion is
met, $C < 1$. The integral in the denominator of the above equation is
equal to $[\Gamma-i\Deltae]^{-1}$ for Lorentzian broadening and equal
to $\sqrt{\frac{\pi}{2}} \frac{w(z_{\mathrm{e}})}{\sigma_{\Delta}}$
with $z_{\mathrm{e}} =
\frac{\Deltae+i\gammaperp}{\sqrt{2}\sigma_{\Delta}}$ for Gaussian
broadening. When the driving is resonant, $\Deltae = 0$, the integral
is equal to $\Gamma^{-1}$ for any (symmetric) distribution according
to Eq.~(\ref{eq:Def_kappac_and_Gamma}).

\begin{figure}[t]
  \centering
  \includegraphics{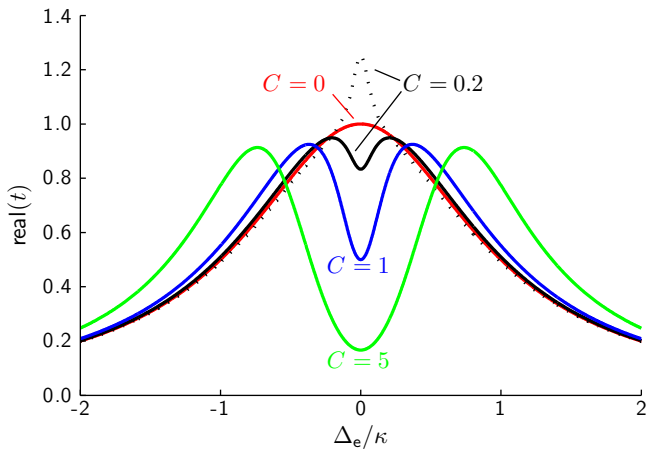}
  \caption{(Color online) The real part of the transmission
    coefficient versus external driving frequency. The calculations
    assume a Lorentzian profile, $\kappa_1 = \kappa_2$, $\kappa =
    10\Gamma$, and $\gens$ is varied in order to obtain the $C$-values
    marked on the curves. Solid lines are non-inverted ($p=-1$) while
    the dotted line for $C=0.2$ corresponds to an inverted sample
    ($p=1$).}
\label{fig:TransmissionSpectrum}
\end{figure}

Now, the reflected and transmitted fields relate to the cavity field
by \cite{Collett.PhysRevA.30.1386(1984)}: $\mean{\aR} =
\sqrt{2\kappa_1}\mean{\ac} - \beta$ and $\mean{\aT} = \sqrt{2\kappa_2}
\mean{\ac}$. This enables a calculation of the complex reflection and
transmission coefficients, $r = \frac{\mean{\aR}}{\beta}$ and $t =
\frac{\mean{\aT}}{\beta}$, respectively. Selected examples have been
plotted in Fig.~\ref{fig:TransmissionSpectrum} for a Lorentzian
inhomogeneous broadening (a Gaussian profile presents qualitatively
similar results). When $C$ increases beyond unity, the bare-cavity
transmission spectrum is significantly modified by the presence of the
spin ensemble and the well-known normal-mode splitting occurs
\cite{Zhu.PhysRevLett.64.2499(1990)}. Note that $C=1$ corresponds to
the case where the transmission coefficient is reduced from unity to
one half (for a symmetric cavity). We also note that the transmission
spectrum exists for an inverted sample when $C < 1$ (the dotted curve
exemplifies this) and that the transmission coefficient may exceed
unity due to the inherent gain of the inverted sample.

A particular relation, which is useful for a simple estimation of the
effective cooperativity parameter, is given by the connection of $C$
to the values of $r$ and $t$ for any (symmetric) distribution on
resonance ($\Deltae = 0$):
\begin{equation}
 pC = \frac{r - \frac{\kappa_1-\kappa_2}{\kappa_1+\kappa_2}}{r+1}
   = 1-\frac{2\sqrt{\kappa_1\kappa_2}}{\kappa}\cdot \frac{1}{t}.
\end{equation}

The fact that the reflection and transmission coefficients may exceed
unity clearly demonstrates that the excitation level of the spin
ensemble must account for the energy balance. This fact is disguised
by the Holstein-Primakoff approximation, but it is possible to
estimate the effect in a mean-field theory on resonance ($\Deltacs =
0$, see the appendix for details):
\begin{equation}
\label{eq:ddt_Sz_main_text}
  \frac{\partial S_z}{\partial t} = -\frac{4p\gens^2|a_{\mathrm{c}}|^2}{\Gamma}.
\end{equation}
We stress that this holds also for a non-inverted sample ($p=-1$), in
which case energy quanta leak from the cavity into the continuous spin
ensemble with a rate proportional to a squared matrix element,
$\sim\frac{\gens^2}{N}$, times the density of states,
$\sim\frac{N}{\Gamma}$, resembling the usual Fermi's-Golden-Rule
expression for decay of a quantum system due to the coupling to a
broad-bandwidth reservoir. In order that this de-polarizing effect is
kept small, the duration $T$ of the external driving must be short
enough so that $\left|\frac{\partial S_z}{\partial t}\right| T \ll
N$. This is equivalent to: $n_{\mathrm{ph}} \ll
\frac{\kappa}{\kappa_1} \cdot\frac{(1-pC)^2}{8|pC|}\cdot N$, where
$n_{\mathrm{ph}} = |\beta|^2 T$ is the total number of photons
supplied by the external driving field during the experiment. Clearly,
for an inverted sample approaching the point of instability, $pC
\rightarrow 1$, the allowed number of photons decreases significantly
below $N$.

\section{Discussion}
\label{sec:Discussion}
The stability criterion of Eq.~(\ref{eq:Stability_condition}) and the
dynamical evolution of the spin-cavity system below and above the
point of stability present the main result of this manuscript. The
results of Sec.~\ref{sec:Evolution_quadratic_moments}, in particular
panels (a-d) of Fig.~\ref{fig:MomentEvolution}, demonstrate that the
stability criterion refers to both the mean values and the second
moments. This follows naturally from the fact that the same matrices
govern the linear sets of equations for the first and second moments,
as shown in the appendix.

Understanding the free evolution of an inverted spin ensemble in a
cavity is of high importance for spin-refocusing techniques. Such
refocusing could improve spin-based quantum memory protocols in
cavities. However, the storage and retrieval part of such a protocol
\cite{Kubo.PhysRevLett.107.220501(2011)} would typically be
implemented in the strong-coupling regime, $\gens \gg \kappa,\Gamma$,
i.e.~with $C\gg 1$, and the ability to tune the value of $C$ during
the experimental protocol would then be necessary. We also note that
the spins can effectively be decoupled from the cavity field by a
large detuning $\Deltacs$. The discussion after
Eq.~(\ref{eq:ac_kick_decay_low_coupling}) can be stated more generally
as $\mean{\ac(\omega)} \approx
\frac{\alpha}{\kappa+i(\Deltacs-\omega)} + \frac{\alpha
  K^+(\omega)}{[\kappa+i\Deltacs]^2}$ when $|\kappa+i\Deltacs|$ is
large compared to the frequency width of $K^+(\omega)$. This equation
reflects the fact that the cavity field follows adiabatically the
evolution of the (effectively) uncoupled spin ensemble (the second
term depends on $\omega$ through $K^+(\omega)$ only), which in turn is
largely given by the Fourier components of $f(\Delta)$ through the
relation~(\ref{eq:Formula_Ktilde_omega}). Note that a broad and smooth
distribution $f(\Delta)$ is required in general, coupled or uncoupled,
if a fast relaxation of both the cavity field and the spin components
is desired.

The diagnostics tools presented in Sec.~\ref{sec:probing} have been
derived for a perfectly polarized spin ensemble ($p=\pm 1$). However,
as exemplified in the appendix by using a suitable sub-ensemble
distribution, we may argue that the results of Sec.~\ref{sec:probing}
hold for a non-perfect polarization also, $-1 \le p \le 1$. The
Holstein-Primakoff approximation corresponds to keeping the collective
spin vector within the linear region around the north or south pole of
a collective Bloch sphere. Relaxing the need for perfect polarization
corresponds to merely reducing the radius of the collective Bloch
vector. We note that the important equations include $p$ and $C$ in
the combination $pC$, and since $C \propto N$ it is reasonable that a
non-perfect spin polarization is accounted for by this
combination. This argument holds also for the stability criterion of
Eq.~(\ref{eq:Stability_condition}).

\section{Conclusion}
\label{sec:Conclusion}
We have demonstrated that inhomogeneous broadening is a stabilizing
mechanism for an inverted spin ensemble coupled to a cavity. A
stability criterion was stated in Eq.~(\ref{eq:Stability_condition}),
and if this criterion is met the transverse spin-component mean values
relax toward zero while the variances of these spin components reach
finite values. This holds simultaneously for the mean values and
variances of the cavity field.

The details of the spin-cavity dynamics was discussed for a Lorentzian
and a Gaussian inhomogeneity in the spin-resonance frequencies. In
particular, the time scale of the relaxation process is well
understood, and fast relaxation requires a broad and smooth
inhomogeneity.
\begin{acknowledgments}
  The authors acknowledge support from the EU integrated project AQUTE
  and the EU 7th Framework Programme collaborative project iQIT. We
  are grateful for useful discussions with C\'ecile Grezes and Patrice
  Bertet.
\end{acknowledgments}

\appendix

\section{Equations for first and second moments using a sub-ensemble
  discretization}
\label{app:Discretization}
The present appendix describes how the Holstein-Primakoff
approximation is applied to establish numerically tractable, linear
equations of motion for the mean values and second moments of field
and collective spin quadrature operators. The system Hilbert space is
infinite dimensional even for a single oscillator mode; however, the
number of equations we need to solve scales only linearly and
quadratically with the number of modes for mean values and second
moments, respectively. In order to handle inhomogeneities numerically,
the ensemble is divided into $M$ sub-ensembles, $\mathcal{M}_1,
\mathcal{M}_2, \ldots, \mathcal{M}_M$, which can each be regarded as
homogeneous with coupling strength $g_m$, spin resonance frequency
$\Delta_m$, and containing $N_m$ spins for $m = 1,\ldots,M$.  We
assume that all spins reside in the ground or in the excited state,
such that $\S_z^{(m)} = \sum_{\mathcal{M}_m}\pauli_z^{(j)} \approx \pm
N_m$, and that the dynamical variables of the spin-cavity system are
described by the operators:
\begin{equation}
  \begin{split}
  \Xa &= \frac{\ac + \adagc}{\sqrt{2}}, \\
  \Pa &= \frac{-i(\ac - \adagc)}{\sqrt{2}}, \\
  \S_x^{(m)} &= \sum_{\mathcal{M}_m}(\pauli_+^{(j)} + \pauli_-^{(j)}), \\
  \S_y^{(m)} &= -i\sum_{\mathcal{M}_m}(\pauli_+^{(j)} - \pauli_-^{(j)}).
  \end{split}
\end{equation}
The $\Xa$ and $\Pa$ operators describe the quadratures of the cavity
field with $[\Xa,\Pa] = i$, while the $\S_k^{(m)}$ components
correspond to twice the total spin in each sub-ensemble with
$[\S_{j}^{(m)},\S_{k}^{(m)}] = 2i\epsilon_{jkl}\S_{l}^{(m)}$.

The strong spin polarization ensures the constant commutator:
$[\S_{x}^{(m)},\S_{y}^{(m)}] = 2i\S_{z}^{(m)} \approx \pm 2 i N_m$,
and validates the ensuing simplified Heisenberg equations of motion in
the Holstein-Primakoff approximation:
\begin{align}
\label{eq:Heis_eq_Xa}
  \frac{\partial \Xa}{\partial t} &= -\kappa \Xa 
    +\Deltacs \Pa  -\sum_m \frac{g_m}{\sqrt{2}}\S_y^{(m)}
    + \hat{F}_{X_{\mathrm{c}}}, \\
\label{eq:Heis_eq_Pa}
  \frac{\partial \Pa}{\partial t} &= -\kappa \Pa 
    -\Deltacs \Xa  -\sum_m\frac{g_m}{\sqrt{2}}\S_x^{(m)}
    + \hat{F}_{P_{\mathrm{c}}}, \\
\label{eq:Heis_eq_Sx_kl}
  \frac{\partial \S_x^{(m)}}{\partial t} &= -\gammaperp \S_x^{(m)}
    -\Delta_m \S_y^{(m)} - \sqrt{2}g_mS_z^{(m)} \Pa + \hat{F}_{S_x^{(m)}}, \\
\label{eq:Mean_val_eq_Sy_kl}
  \frac{\partial \S_y^{(m)}}{\partial t} &= -\gammaperp \S_y^{(m)}
    +\Delta_m \S_x^{(m)} - \sqrt{2}g_mS_z^{(m)} \Xa + \hat{F}_{S_y^{(m)}}.
\end{align}
Note, the external driving is assumed to be absent, $\beta = 0$. The
last term in each equation is a Langevin noise operator, the
properties of which follow from the quantum Langevin equations of a
damped harmonic oscillator \cite{Gardiner.QuantumNoise}. For instance,
the preservation of commutators require that
$[\hat{F}_{X_{\mathrm{c}}}(t), \hat{F}_{P_{\mathrm{c}}}(t')] = 2\kappa
\cdot i\delta(t-t')$ and $[\hat{F}_{S_x^{(m)}}(t),
\hat{F}_{S_y^{(m)}}(t')] = 4\gammaperp N_m\cdot i\delta(t-t')$.

Arranging the field and spin operators in a column vector
$\hat{\vec{y}}$ with $2M+2$ components, we can write the coupled
Heisenberg equations of motion in the compact form:
\begin{equation}
\label{Heis-y}
  \frac{\partial}{\partial t}\hat{\vec{y}}
    = \mat{M}\hat{\vec{y}} + \hat{\vec{F}},
\end{equation}
where the driving matrix $\mat{M}$ is given by:
\begin{equation}
  \mat{M} =
  \begin{bmatrix}
    \mat{A} & \mat{B}^{(1)} & \mat{B}^{(2)} & \ldots & \mat{B}^{(M)} \\
    \mat{C}^{(1)} & \mat{D}^{(1)} & 0 & \ldots & 0 \\
    \mat{C}^{(2)} & 0 & \mat{D}^{(2)} & \ldots & 0 \\
    \vdots & \vdots & \vdots & \ddots & \vdots \\
    \mat{C}^{(M)} & 0 & 0 & \ldots & \mat{D}^{(M)}
  \end{bmatrix},
\end{equation}
with
\begin{equation}
  \begin{split}
  \mat{A} &=
  \begin{bmatrix}
    -\kappa &  \Deltacs \\ -\Deltacs & -\kappa
  \end{bmatrix}, \quad
  \mat{B}^{(m)} =
  \begin{bmatrix}
    0 & -\frac{g_m}{\sqrt{2}} \\ -\frac{g_m}{\sqrt{2}} & 0
  \end{bmatrix}, \\
  &\mat{C}^{(m)} =
  \begin{bmatrix}
    0 & -\sqrt{2}g_m S_z^{(m)} \\
    -\sqrt{2}g_m S_z^{(m)} & 0
  \end{bmatrix}, \\
  &\mat{D}^{(m)} =
  \begin{bmatrix}
    -\gammaperp & -\Delta_m \\
    \Delta_m & -\gammaperp
  \end{bmatrix}.
  \end{split}
\end{equation}
Inserting $\hat{\vec{y}}=\vec{y} + \delta\hat{\vec{y}}$, where
$\mean{\delta\hat{\vec{y}}} = 0$, into Eq.~(\ref{Heis-y}) yields the mean
value equation for $\vec{y}$:
\begin{equation}
\label{eq:dydt=My}
  \frac{\partial\vec{y}}{\partial t} = \mat{M}\vec{y}.
\end{equation}
This is the equation solved in our numerical mean field analysis, and
it is the eigenvalues of the matrix $\mat{M}$ which govern the
stability of the solutions. The Heisenberg equations for the
fluctuations around the mean values
\begin{equation}
  \frac{\partial}{\partial t}\delta\hat{\vec{y}}
     = \mat{M}\delta\hat{\vec{y}} + \hat{\vec{F}},
\end{equation}
are operator valued, and to investigate the fluctuations numerically
we introduce the $(2+2M)\times(2+2M)$ covariance matrix $\CorrMat =
2\Re\{\mean{\delta\hat{\vec{y}} \cdot
  \delta\hat{\vec{y}}^{\mathrm{T}}}\}$ with elements $\CorrMat_{kl} =
C(\hat{y}_k,\hat{y}_l) = 2\Re\{\mean{\delta\hat{y}_k
  \delta\hat{y}_l}\}$, which represent the quantum correlations
between any two of the $2+2M$ relevant operators, $\hat{y}_k$ and
$\hat{y}_l$. In particular $\CorrMat_{kk} = 2\Var(\hat{y}_k)$. The
special form of $\CorrMat$ leads to its time derivative:
\begin{equation}
\label{eq:ddt_CorrMat}
  \frac{\partial\CorrMat}{\partial t} = \mat{M}\CorrMat
     +\CorrMat\mat{M}^{\mathrm{T}} + \mat{N},
\end{equation}
where $\mat{N}$ is related to the correlation of the Langevin
operators by $\mat{N}\delta(t-t') =
2\Re\{\mean{\hat{\vec{F}}(t)\hat{\vec{F}}(t')^T}\}$, which for
reservoirs at zero temperature amounts to:
\begin{equation}
\label{eq:def_noise_matrix}
  \begin{split}
  \mat{N} &=
  \begin{bmatrix}
    \mat{V} & 0 & \ldots & 0 \\
    0 & \mat{U}^{(1)} & \ldots & 0 \\
    \vdots & \vdots & \ddots & \vdots \\
    0 & 0 & \ldots & \mat{U}^{(M)}
  \end{bmatrix}, \quad
  \mat{V} =
  \begin{bmatrix}
    2\kappa & 0 \\ 0 & 2\kappa
  \end{bmatrix}, \\
  &\mat{U}^{(m)} =
  \begin{bmatrix}
    4\gammaperp N_m & 0 \\
    0 & 4\gammaperp N_m
  \end{bmatrix}.
  \end{split}
\end{equation}
We observe that the eigenvalue spectrum of the matrix $\mat{M}$ also
accounts for the stability properties of the covariance matrix, and
thus the second moments of collective spin variables.

The sub-ensemble grouping of spins serves two purposes. Most
importantly, it enables the application of the Holstein-Primakoff
approximation, which results in linear coupled equations for the first
and second moments of effective oscillator quadrature operators. This
significantly reduces the number of dynamical variables accounting for
the full quantum state to $(2M+2)$ mean values and $(2M+2)^2$ second
moments with $M$ being the number sub-ensembles. For the validity and
accuracy of our approach one should ensure that $M$ is large enough to
adequately represent the inhomogeneous broadening of the spin
ensemble, i.e., the frequency spacing must be sufficiently small to
avoid discretization errors such as artificial revivals of the spin
state, while still treating a sufficiently large number of spins to
render the Holstein-Primakoff oscillator description valid.

Although we treat the collective $\S_z^{(m)}$ operators as constants
equal to $\pm N_m$, the validity of the Holstein-Primakoff
approximations merely relies on their mean values being much larger
than their quantum fluctuations. Our analysis will thus also apply for
partly polarized samples, and we may revisit the full Heisenberg
equations of motion in order to determine if they change in time due
to the coupling to the cavity field. To this end, consider the time
derivative of $S_z^{(m)} = \mean{\S_z^{(m)}}$:
\begin{equation}
\label{eq:ddt_Szm_v1}
  \frac{\partial S_z^{(m)}}{\partial t} = -2ig_m(\mean{\S_+^{(m)}\ac}
    -\mean{\S_-^{(m)}\adagc}).
\end{equation}
In the case of a seeded cavity, studied in Sec.~\ref{sec:probing}, the
spin and field operators have finite mean values, and the above
expectation values approximately factor: $\mean{\S_+^{(m)}\ac} \approx
S_+^{(m)}a_{\mathrm{c}}$, etc. From the mean value equation
\begin{equation}
  \frac{\partial S_-^{(m)}}{\partial t} = -(\gammaperp +i\Delta_m) S_-^{(m)}
  +ig_m S_z^{(m)}a_{\mathrm{c}},
\end{equation}
we can adiabatically eliminate the spin variable:
$S_-^{(m)} = \frac{ig_m S_z^{(m)}a_{\mathrm{c}}}
{\gammaperp+i\Delta_m}$ to a good approximation when
$a_{\mathrm{c}}$ varies slowly ($\Deltacs = 0$). Setting $S_z^{(m)}
= p N_m$ in this expression, and inserting the result into
Eq.~(\ref{eq:ddt_Szm_v1}), the change of $S_z$ becomes:
\begin{equation}
  \begin{split}
  \frac{\partial S_z}{\partial t} &= \sum_m \frac{\partial S_z^{(m)}}{\partial t}
   = -4p|a_{\mathrm{c}}|^2\sum_m \frac{g_m^2 N_m}{\gammaperp + i\Delta_m} \\
   &= -4p\gens^2|a_{\mathrm{c}}|^2\int_{-\infty}^{\infty} \frac{f(\Delta)d\Delta}
      {\gammaperp + i\Delta}.
  \end{split}
\end{equation}
The second equality is valid for a symmetric sub-ensemble
distribution, while the third equality assumes the continuum limit of
the sub-ensemble description of the actual inhomogeneous
distribution. Using Eq.~(\ref{eq:Def_kappac_and_Gamma}) leads to the
Fermi's-Golden-Rule-like expression~(\ref{eq:ddt_Sz_main_text}), which
both explains the field loss due to absorption by the non-inverted
spin ensemble and the gain obtained due to stimulated emission by the
inverted sample. We note that in the absence of coherent driving, mean
values of the cavity field and the spin components vanish in steady
state, and the product term in Eq.~(\ref{eq:ddt_Szm_v1}) is a
combination of the second moments obtained by solving
Eq.~(\ref{eq:ddt_CorrMat}). For the special case of an inverted
homogeneous or Lorentzian spin ensemble, these second moments are
given by Eq.~(\ref{eq:Variances_SS}), and the rate of change in $S_z$
becomes $\frac{\partial S_z}{\partial t} =
-\frac{4\gens^2}{(\kappa+\Gamma)(1-C)}$ when $p=1$ and zero when
$p=-1$.


%

\end{document}